\documentclass{article}
\usepackage{subcaption}
\usepackage{multirow}
\usepackage{nicefrac}
\usepackage{booktabs}
\usepackage{stfloats}
\usepackage{colortbl}
\usepackage[most]{tcolorbox}
\usepackage{paralist}
\usepackage{xspace}
\usepackage{mdframed}

\usepackage{caption} 
\usepackage{url}
\newcommand{\Atwelve}{\ensuremath{\hat{A}}\textsubscript{12}\xspace}
\newcommand{\pearsonCoeff}{\ensuremath{\mathit{R}}\xspace}
\newcommand{\determCoeff}{\ensuremath{\mathit{R^2}}\xspace}
\newcommand{\RMSRE}{\ensuremath{\mathit{RMSRE}}\xspace}
\newcommand{\RMSE}{\ensuremath{\mathit{RMSE}}\xspace}
\newcommand{\MAPE}{\ensuremath{\mathit{MAPE}}\xspace}
\newcommand{\TestLoss}{\ensuremath{\mathit{TestLoss}}\xspace}
\newcommand{\odds}{\ensuremath{\mathit{odds}}\xspace}

\newcommand{\oddsRatio}{\ensuremath{\mathit{Odr}}\xspace}
\newcommand{\DPE}{\ensuremath{\mathit{Dpe}}\xspace}
\newcommand{\stateweight}{\ensuremath{\mathit{Stw}}\xspace}
\newcommand{\circuitwidth}{\ensuremath{\mathit{Cw}}\xspace}
\newcommand{\circuitdepth}{\ensuremath{\mathit{Cd}}\xspace}
\newcommand{\oneqgate}{\ensuremath{\mathit{Gc_{1q}}}\xspace}
\newcommand{\twoqgate}{\ensuremath{\mathit{Gc_{2q}}}\xspace}
\newcommand{\prob}{\ensuremath{\mathit{Prob_{obv}}}\xspace}
\newcommand{\dpeone}{\ensuremath{\mathit{Dpe_{\nicefrac{1}{4}}}}\xspace}
\newcommand{\dpetwo}{\ensuremath{\mathit{Dpe_{\nicefrac{1}{2}}}}\xspace}
\newcommand{\dpethree}{\ensuremath{\mathit{Dpe_{\nicefrac{3}{4}}}}\xspace}

\newcommand{\QRAFT}{QRAFT\xspace}
\newcommand{\ourApproach}{Q-LEAR\xspace}
\newcommand{\groundstate}{\textit{GS}\xspace}
\newcommand{\pricingcall}{\textit{PC}\xspace}
\newcommand{\pricingput}{\textit{PP}\xspace}
\newcommand{\qaoa}{\textit{QAOA}\xspace}
\newcommand{\routing}{\textit{RT}\xspace}
\newcommand{\tsp}{\textit{TSP}\xspace}
\newcommand{\outputstateerror}{\textit{Output State Error}\xspace}
\newcommand{\outputerror}{\textit{Output Error}\xspace}

\usepackage{PRIMEarxiv}

\title{A Machine Learning-Based Error Mitigation Approach for Reliable Software Development on IBM’s Quantum Computers}

\author{
  Asmar~Muqeet\\
  Simula Research Laboratory \\
  University of Oslo \\
  Oslo\\
  \texttt{asmar@simula.no} \\
   \And
  Shaukat~Ali \\
  Simula Research Laboratory and \\
  Oslo Metropolitan University \\
  Oslo\\
  \texttt{shaukat@simula.no} \\
   \And
  Tao~Yue \\
  Simula Research Laboratory \\
  Oslo\\
  \texttt{taoyue@gmail.com} \\
   \And
  Paolo~Arcaini \\
  National Institute of Informatics \\
  Tokyo\\
  \texttt{arcaini@nii.ac.jp} \\
}

\begin{document}
\maketitle

\begin{abstract}
Quantum computers have the potential to outperform classical computers for some complex computational problems. However, current quantum computers (e.g., from IBM and Google) have inherent noise that results in errors in the outputs of quantum software executing on the quantum computers, affecting the reliability of quantum software development. The industry is increasingly interested in machine learning (ML)-based error mitigation techniques, given their scalability and practicality. However, existing ML-based techniques have limitations, such as only targeting specific noise types or specific quantum circuits. This paper proposes a practical ML-based approach, called \ourApproach, with a novel feature set, to mitigate noise errors in quantum software outputs. We evaluated \ourApproach on eight quantum computers and their corresponding noisy simulators, all from IBM, and compared \ourApproach with a state-of-the-art ML-based approach taken as baseline. Results show that, compared to the baseline, \ourApproach achieved a 25\% average improvement in error mitigation on both real quantum computers and simulators. We also discuss the implications and practicality of \ourApproach, which, we believe, is valuable for practitioners.
\end{abstract}


\section{Introduction}
Quantum Computing (QC) holds immense promise for tackling complex computational problems beyond the capabilities of classical computers~\cite{speed}. However, the practical realization of this potential faces challenges, with quantum noise being a prominent obstacle. \textit{Quantum noise}, stemming from imperfections and environmental interactions, significantly impacts the accuracy of computations performed by quantum computers~\cite{noise_source}. Consequently, the accuracy of software\footnote{Quantum software is currently being built as quantum circuits, i.e., a sequence of quantum gate operations applied to quantum bits (qubits).} running on a noisy quantum computer is compromised, even when correctly implemented, thereby limiting QC's practical applications and advantage over classical computers. 

Developing quantum software while addressing noise poses various challenges, including uncertainty about whether the quantum software is producing an incorrect output or if the output is flawed due to noise in the quantum computer. Recognizing the noise issue, industry leaders in QC, such as IBM, have identified \textit{quantum error correction} (i.e., error correction during circuit executions) and \textit{quantum error mitigation} (i.e., error correction post-circuit execution) as pivotal building blocks in their roadmap to facilitate the development of practical QC software~\cite{levels,raodmap}. This paper focuses on quantum error mitigation, i.e., applying automated error mitigation techniques after software execution on a quantum computer to eliminate the noise effects from the software outputs. Such noise elimination serves as a valuable tool for quantum software engineers, which facilitates software development and testing with outputs that have undergone a noise-cleansing process. Doing so, thereby, increases software engineers' assurance of the correctness of quantum software under real-world quantum computing conditions--inherent noise in quantum computers.

In practice, several error mitigation techniques have been incorporated into industrial frameworks such as IBM's Qiskit~\cite{qiskit} to correct output errors in quantum circuits. Notable methods include Probabilistic Error Cancellation (PEC)~\cite{pec1} and Zero-Noise Extrapolation (ZNE)~\cite{zne}. While these techniques show promise in mitigating output errors, they often require a comprehensive understanding of specific noise characteristics for each circuit. For instance, to use PEC, it is needed to identify the predominant type of noise error and create mathematical models for each type of noise that can impact a given circuit. However, this process incurs an exponential cost in terms of circuit sampling (i.e., the number of repeated executions of a circuit required to build a noise model), rendering it impractical in terms of scalability for current quantum computers~\cite{peccost}. On the other hand, ZNE is accurate only for specific circuits where noise lacks temporal correlation, a condition not met by the majority of current quantum algorithms~\cite{zne_limit}.

In recent years, there has been a shift among industry practitioners towards machine learning (ML)-based error mitigation for practical, reliable, and scalable solutions~\cite{mlforQem}, with state-of-art being \QRAFT~\cite{qraft} which leverages an ensemble-based ML algorithm for quantum error mitigation in the presence of noise. However, a critical limitation of the current methods, including \QRAFT, is the absence of a reliable feature set that can accurately quantify the noise magnitude of a quantum circuit. Consequently, the ML models of these methods could exacerbate errors (instead of removing them) by making inaccurate adjustments based on wrong noise estimates, as evidenced by the results of our empirical study.

This paper presents an ML-based error mitigation approach (\ourApproach) to address the limitations of current ML-based error mitigation methods. \ourApproach proposes a set of novel features, including the Depth-cut Program Error (\DPE), which cuts a quantum circuit at specific circuit depths and leverages quantum operations' reversibility feature to quantify noise magnitude. With \DPE, we estimate noise magnitude more accurately when compared with the state of the art. \ourApproach enables ML models to effectively learn and mitigate quantum circuit output error caused by noise. We empirically evaluate the effectiveness of \ourApproach with various ML models on real quantum circuits and across eight IBM's quantum computers and their corresponding noisy simulators. Results show that ML models trained with \ourApproach perform significantly better than \QRAFT on IBM's quantum computers and simulators. Notably, \ourApproach demonstrates an average improvement of 25\% in error mitigation compared to \QRAFT when executed on eight IBM's quantum computers and simulators. The results emphasize that \ourApproach's ML model trained with its feature set has the potential to substantially improve the reliability of quantum software development, especially on IBM's quantum computers. 

\section{Background}\label{sec:background}

\subsubsection*{\bf Qubit}
Quantum Computing (QC) uses {\it quantum bits}, i.e., {\it qubits}, as its fundamental information units. 
A qubit can exist in a superposition of states $|0\rangle$ and $|1\rangle$, with associated amplitudes for each, and the state of one qubit immediately influences the state of another one when they are entangled. The amplitude is a complex number comprising both magnitude and phase in its polar representation. In the Dirac notation~\cite{dirac}, a qubit is denoted as $|\psi\rangle = \alpha |0\rangle + \beta |1\rangle$, where $\alpha$ and $\beta$ represent the amplitudes associated with states $|0\rangle$ and $|1\rangle$, respectively. The superposition of a qubit upon measurement collapses to a single basis state where the probability of observing a qubit in either state $|0\rangle$ or $|1\rangle$ is determined by the squared magnitude of $\alpha$ and $\beta$, with the sum of all squared magnitudes being 1: $|\alpha|^2 + |\beta|^2 = 1$.
\subsubsection*{\bf Quantum Gate and Circuit}
\begin{figure}[!tb]
\centering
\begin{subfigure}[t]{0.4\columnwidth}
\centering
\includegraphics[width=0.95\textwidth]{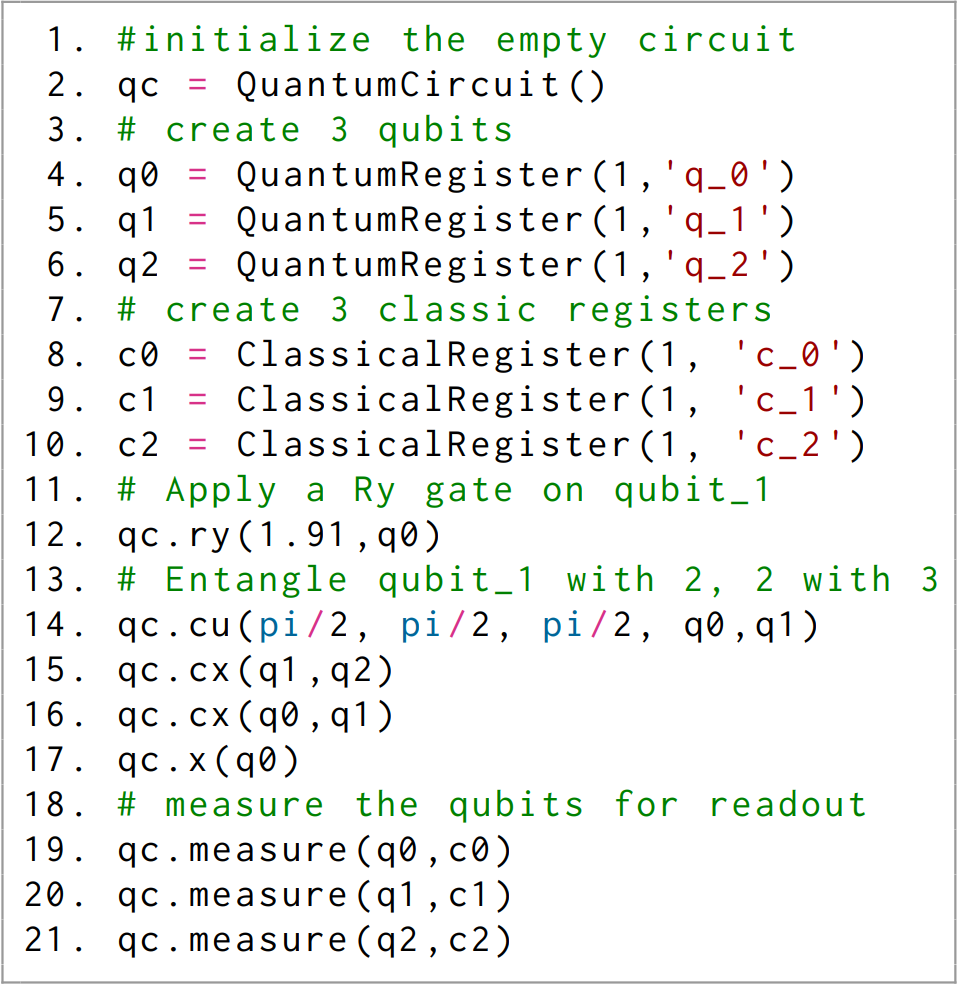}
\caption{Python code in Qiskit}
\label{fig:wstate_python}
\end{subfigure}
\begin{subfigure}[t]
{0.4\columnwidth}
\centering
\includegraphics[width=0.95\textwidth]{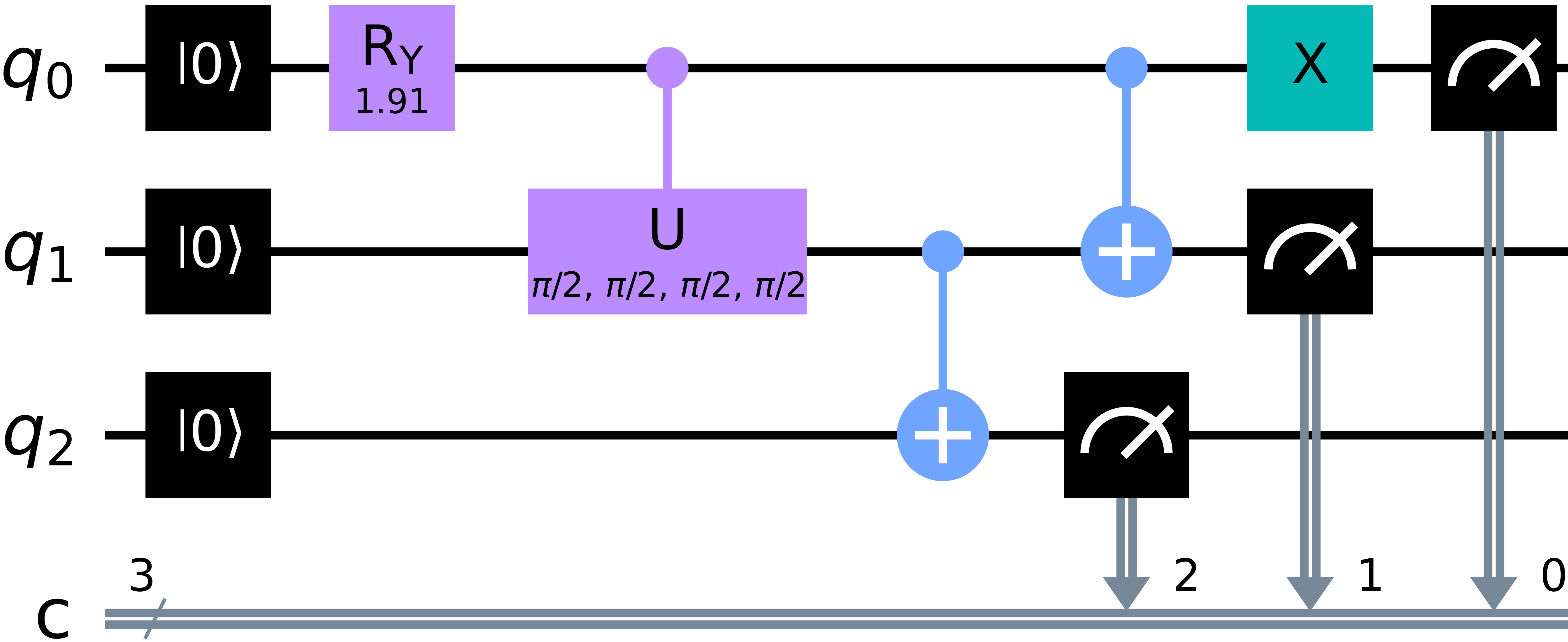}
\caption{Quantum Circuit}
\label{fig:wstate_circuit}
\end{subfigure}

\caption{W-state quantum program}
\label{fig:wstate}
\end{figure}

Gate-based quantum computers manipulate a qubit with a quantum gate, which is a unitary operator that changes the qubit's state based on a unitary matrix~\cite{basic}. For example, a \textit{Hadamard} gate puts a single qubit in superposition. Currently, quantum programs are represented in the circuit model~\cite{circuitmodel}; a quantum program is a sequence of quantum gates acting on a set of qubits. Each gate operation is a single-time step in the unitary evolution of a quantum system~\cite{circuitmodel}. \textbf{Circuit depth} is an essential indicator of circuit complexity, representing the longest sequence of gate operations in a circuit. Fig.~\ref{fig:wstate_python} shows the Python code for the program that creates a three-qubit entangled state known as W-state~\cite{wstateurl} and Fig.~\ref{fig:wstate_circuit} shows its corresponding quantum circuit.
%
%
%
In Fig.~\ref{fig:wstate_python}, lines 1-6 create an empty quantum circuit with three qubits $q_0,q_1,q_2$ initialized to state $|0\rangle$. Then, lines 7-10 create three classical registers $c_0,c_1,c_2$ to hold the measurement results of the qubits. In line 12, the $R_y$ gate operation is performed on the circuit to rotate the Y axis of $q_0$'s Bloch sphere by 1.91 degrees. Next, in line 14 a conditional {\it Hadamard} gate is applied on $q_1$ to put it in superposition. In lines 15 and 16, two conditional \textit{NOT} (\textit{CNOT}) gates are then applied to entangle all three qubits. After applying another \textit{NOT} gate (also called \textit{X} gate) in line 17, we obtain a W state ($|W\rangle = \frac{1}{\sqrt{3}}(|001\rangle + |010\rangle + |100\rangle)$. Lines 19-21 apply the measurement operation on all three qubits so that we can get as output one of the three states ($|001\rangle, |010\rangle, |100\rangle$) with equal probability on the classical registers $c_0$, $c_1$, and $c_2$.

\subsubsection*{\bf Transpilation}
Each quantum computer has its own native gate set and qubit connection topology, implying that each qubit can only interact with another qubit if a physical connection exists between them. Hence, some gate operations cannot be performed due to limited physical connections. {\it Transpilation} transforms a quantum logical circuit to a transpiled circuit only containing hardware-defined gate operations and additional swap operations for solving the limited physical connections of the hardware~\cite{circuitmodel}.

\subsubsection*{\bf Quantum Noise}
{\it Noise} arises from various sources. First, environmental factors, e.g., magnetic fields and radiation can impact computations~\cite{Chuang1995, noise_source}. When qubits interact with their surroundings, these interactions can cause disturbances and information loss in quantum states, called {\it decoherence}~\cite{decoherence_def}. Second, even when qubits are isolated from the environment, unwanted interactions can occur among them, resulting in {\it crosstalk noise}~\cite{decoherence1,decoherence2,decoherence3}. Third, {\it imprecise calibrations} of quantum gates, which are necessary to optimize gate parameters and reduce errors while improving fidelity, can introduce noise~\cite{callibration}. Small calibration errors may cause minor changes in qubit phases, amplitudes, etc., resulting in undesired states after a series of gate operations~\cite{gatenoise}.

Noise in computing systems is not a unique concept to quantum computing; it also exists in the classical world, notably in domains like the Internet of Things (IoT) and cyber-physical systems~\cite{sensornoise, iotnoise}. This raises the question of whether classical noise filtering or error correction techniques can be directly applied to quantum computing (QC). While some principles from classical methods, such as error correction codes derived from information theory, can find applications in QC~\cite{qec1, qec2}, it is crucial to acknowledge that quantum noise possesses unique characteristics. Quantum noise exhibits phenomena such as entanglement, superposition, and quantum interference, which differentiate it from classical noise. These quantum characteristics make the treatment of quantum noise significantly more complex. In contrast, classical noise can be described using classical probability theory and arises from random fluctuations, electronic interference, thermal effects, etc.~\cite{classicalnoise}. Classical noise sources often exhibit behaviors where noise events are independent and adhere to well-understood probability distributions like Gaussian or Poisson distributions~\cite{classicalnoise}. However, when it comes to quantum noise, understanding the underlying distribution becomes extremely challenging. This difficulty arises primarily due to the restrictions in quantum computing inherent to quantum mechanics, such as the no-cloning theorem and state collapse~\cite{nocloning}. These quantum principles fundamentally limit the ability to clone quantum states and introduce uncertainties in the measurement process, making it challenging to model quantum noise with the same level of predictability as classical noise. 

\section{Related Work and Their Limitations}
Existing quantum noise error mitigation methods can be classified into three categories.

\textbf{Probabilistic Error Cancellation (PEC).}
Since being introduced in~\cite{pec1} for Markovian noise (errors at a time point independent of what occurred in the past), PEC has been extended to handle non-Markovian noise by~\cite{pec2}. It uses the quasi-probability decomposition of the inverse noise process, resulting in a linear combination of noisy circuits. Various PEC methods have recently been proposed, including~\cite{pec3,pec4,pec5}, for various noise errors.
\textit{\textbf{Limitation: }}However, PEC-based methods require complete knowledge about noise characteristics specific to a circuit, including identifying the dominant noise error type and establishing mathematical models for each noise channel in the circuit. Hence, applying PEC to different quantum circuits executing on different quantum computers becomes extremely challenging.

\textbf{Zero-Noise Extrapolation (ZNE).}
ZNE gathers execution data of quantum circuits at different error rates and extrapolates to the zero noise limit~\cite{zne}. Various studies have extended ZNE with different extrapolation methods~\cite{zne,pec1,zne2,zne3}. \textit{\textbf{Limitation: }}However, ZNE-based methods assume that noise is uncorrelated with time, which has been invalidated by recent studies~\cite{zne_limit}. In the presence of time-correlated noise, scaling quantum circuits for different error rates without altering their spectral distribution becomes difficult~\cite{zne_limit}.

\textbf{Learning-based methods.} Recent research focuses on using ML for noise error mitigation to deal with some limitations of methods such as PEC and ZNE. Examples include Clifford data regression (CDR)~\cite{cdr}, Learning-based PEC (L-PEC)~\cite{lpec}, and \QRAFT~\cite{qraft}. CDR employs near-Clifford quantum circuits as training data to develop a regression model that mitigates noise; however, it works only for quantum circuits with Clifford gates. L-PEC does not rely on prior knowledge of noise channels as other PEC methods do. It generates multiple variants of the target quantum circuit by replacing non-Clifford gates with gates that are easier to simulate classically. The execution of these variants is then used as training data for a probabilistic noise mitigation model. However, L-PEC assumes that single-qubit Clifford gates are noise-free, which is not always true for real-world quantum computers~\cite{singlequbitgate}. 

\QRAFT 
leverages the reversibility of quantum circuits as a pseudo oracle for training a noise mitigation regression model. \textbf{\textit{Limitation: }}Although \QRAFT avoids assumptions about noise and quantum gate types, it doubles the circuit depth for feature calculation, so increasing the decoherence likelihood and making error mitigation difficult~\cite{depthimpact}. To this end, we introduce a novel Depth-cut Program Error~(\DPE) feature in \ourApproach that uses the same pseudo oracle as \QRAFT, but divides a quantum circuit into smaller subcircuits to avoid doubling the original circuit depth~(see section~\ref{sec:outLevelFeatures}). Further, \QRAFT is more prone to cross-talk noise in feature calculations, as it uses all qubits in the reversed quantum circuit, leading to additional cross-talk noise compared to the original circuit and potentially resulting in inaccurate noise magnitude estimations. In contrast, we only measure a subset of qubits for our \DPE feature that are measured in the original circuit.

\section{Preliminaries}
We describe the following terms to ease understanding of the paper.

\paragraph{Output State}
It is one of the possible states observed after the quantum circuit execution. For instance, in the 3-qubit W-state quantum circuit (Fig.~\ref{fig:wstate}), there are $2^3$ (i.e., 8) possible output states (see Table~\ref{tab:wstate}).
\begin{table}[!tb]
\caption{Ideal and noisy outputs of the W-state circuit (Fig.~\ref{fig:wstate}) after 1024 executions on IBM's Quito (5-qubit quantum computer). Column \textit{Output States with Probabilities} shows \textbf{Output States} with associated probabilities
}
\label{tab:wstate}
\centering
\resizebox{0.6\columnwidth}{!}{%
\begin{tabular}{c|cccccccc}
\toprule
\textbf{} & \multicolumn{8}{c}{\textbf{Output States with Probabilities}} \\
\midrule
\textbf{Circuit Output} & \textbf{000} & \textbf{001} & \textbf{010} & \textbf{011} & \textbf{100} & \textbf{101} & \textbf{110} & \textbf{111} \\
\midrule
\textbf{Ideal} & 0 & 0.33 & 0.33 & 0 & 0.33 & 0 & 0 & 0 \\
\textbf{Noisy} & 0.002 & 0.034 & 0.29 & 0.007 & 0.29 & 0.011 & 0.009 & 0.004 \\
\bottomrule
\end{tabular}
}
\end{table}
Each output state has a probability determined by the circuit's logic and operations.

\paragraph{Circuit Output}
It refers to all output states and their respective probabilities observed after circuit execution. For example, the ideal output of W-state (row 1 of Fig.~\ref{fig:wstate}) comprises 8 possible output states and their probabilities.

\paragraph{Output Error}
Quantum noise manifests as errors in the quantum circuit's output. For example, the \textit{Noisy} row of Table~\ref{tab:wstate} shows the noisy outputs of the W-state circuit (see Fig.~\ref{fig:wstate}), where we observe wrong output states and probabilities compared to \textit{Ideal}. In literature, to quantify the amount of noise in the circuit output, Hellinger distance~(HLD) has been used to calculate the similarity between ideal and observed circuit outputs~\cite{Hellinger1, Hellinger2, hellinger3}: 
\begin{equation}\label{eq:HL_X}
h(P,Q) = \frac{1}{\sqrt{2}}||\sqrt{P}-\sqrt{Q}||_2
\end{equation}
where $P$ and $Q$ are the true and observed noisy probability distributions of a circuit's outputs. Hellinger distance~(HLD) is between 0 and 1, where 0 means that two outputs are identical and 1 is the opposite. For example, the HLD between the \textit{Ideal} and \textit{Noisy} outputs in Table~\ref{tab:wstate} is $0.2$, i.e., a 20\% difference in the output.

\paragraph{Output State Error}
\begin{equation}\label{eq:stateerror}
\mathit{err}_s = |P_s-Q_s|
\end{equation}
where $P_s$ is the ideal probability, and $Q_s$ is the observed noisy probability of a state. For instance, the output state error for $|001\rangle$ (Table~\ref{tab:wstate}) is $|0.33-0.034| = 0.296$.

\section{Approach}
\begin{figure*}[!tb]
\centering
\includegraphics[width=\textwidth]{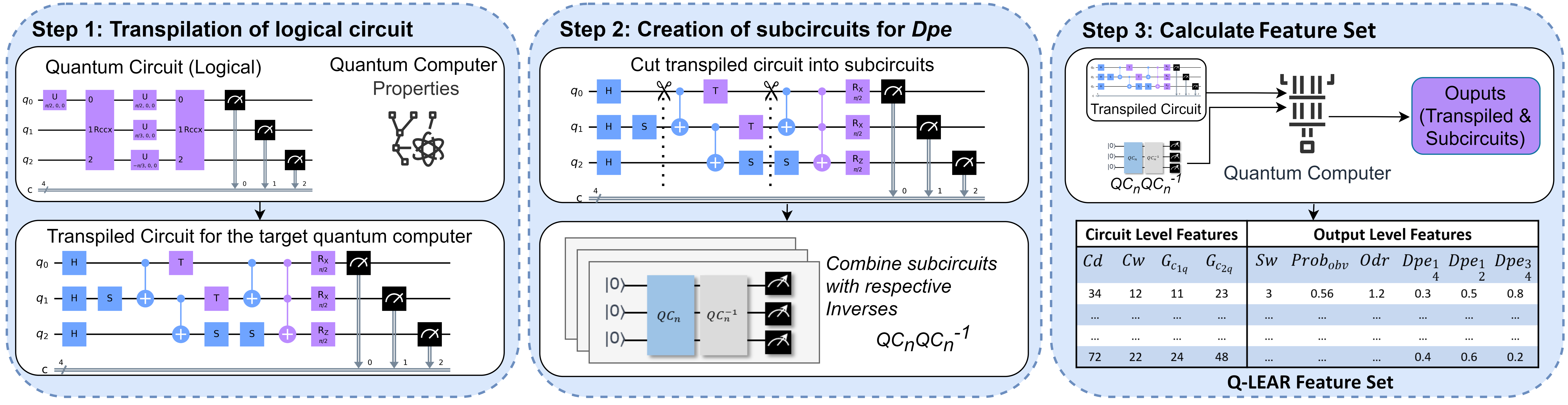}
\centering
\caption{Overview of the process of calculating \ourApproach's feature set for a given quantum circuit and quantum computer}
\label{fig:overview}
\end{figure*}

We propose a \textbf{L}earning-based \textbf{E}rror mitigation \textbf{A}pproach with a \textbf{R}obust feature set (Quantum-LEAR or \ourApproach for short), to reduce the effect of noise from program output. \ourApproach has circuit- and output-level features. Fig.~\ref{fig:overview} shows the process (having three steps) for calculating the \ourApproach features for a given quantum circuit and a target quantum computer.
First, we transpile a quantum circuit (see Sect.~\ref{sec:background}). Second, we divide the transpiled circuit into subcircuits for \DPE feature calculation (detailed in Sect.~\ref{sec:outLevelFeatures}). Third, we execute the transpiled circuit and all subcircuits on the quantum computer or simulator and calculate \ourApproach's feature set for each output state derived from the circuit and its output.

\subsection{Circuit-level Features} \label{subsec:staticfeatures}
Circuit-level features are:
\begin{inparaenum}[1)]
\item the total number of qubits, also called \textbf{circuit width} (\circuitwidth),
\item the \textbf{circuit depth} (\circuitdepth),
\item the number of \textbf{single gate operations} (\oneqgate), and
\item the number of \textbf{two-qubit gate operations} (\twoqgate).
\end{inparaenum}
We chose \circuitwidth since its higher value leads to a higher probability of cross-talk noise, hence inducing errors in the circuit output~\cite{crosstalk}. \circuitdepth also directly relates to the effect of noise on the circuit output~\cite{depthimpact} as decoherence increases with the increased depth.
Quantum gates also play a crucial role in characterizing noise effect since different numbers of single and two-qubit gates are impacted by noise in distinct ways~\cite{gatediff}.

Note that after transpilation, the circuit-level features, e.g., depth, can significantly differ from the logical circuit. Thus, we calculated features on the transpiled circuits. In our feature set related to gate operations in a quantum circuit, we specifically focus on generic features that can be calculated for gate-based quantum computers. This approach involves excluding non-generic gate features like counts of U1, U2, and U3 gates, which were used in \QRAFT. In transpiled circuits, U1, U2, and U3 gates are substituted with varying numbers of other single and two-qubit gates. The choice of replacement gates depends on the logic of the quantum algorithm and the gate set supported by the target hardware (i.e., the one and two-qubit gates physically available in the quantum hardware). While different quantum computers may have distinct gate sets, the commonality is that presently only one and two-qubit gates are physically supported. To ensure compatibility across various quantum computers, we have chosen to categorize gates based on the number of qubits they act on in the transpiled circuit.

\subsection{Output-level Features}\label{sec:outLevelFeatures}
From a quantum circuit output, we use the following output-level features.
\textbf{Observed Probability}~(\prob) is the probability associated with an output state observed after execution on the quantum computer or simulator.
%
\textbf{Odds Ratio}~(\oddsRatio) quantifies the strength of association between each output state of two consecutive quantum circuit executions, defined as $\oddsRatio = \frac{\odds_r}{\odds_{r+1}}$, where $\odds_r$ and $\odds_{r+1}$ are the odds of a specific output state in two consecutive executions~\cite{odds}. An event's odd is calculated as $\odds_r = \frac{P_s}{1-P_s}$, where $P_s$ is the probability of the output state \textit{s} in one circuit execution. Compared to probabilities of output states, the odds ratio offers a slight advantage under noise due to its scale-invariance property, i.e., remaining unchanged even when all probabilities are multiplied by a constant noise factor~\cite{odds}. This property makes the odds ratio less susceptible to constant noise factors~\cite{odds}.

\textbf{State Weight}~(\stateweight) is the number of qubits in state $|1\rangle$. It was first studied by~\cite{qraft} as a feature for mitigating noise error; the study shows that states with lower weights have higher noise errors as higher-weight states (with more qubits in $|1\rangle$) have a higher chance of relaxing to states with lower weights due to noise. Relaxed qubits end up accumulating errors in lower-weight states.

\textbf{Depth-cut Program Error~(\DPE)} measures the noise magnitude affecting the circuit's output. \QRAFT~\cite{qraft} uses the inverse of a quantum circuit to quantify noise impact on the circuit output. However, this doubles the circuit depth, increasing decoherence noise and weakening correlation between noise effects on inverted and original circuits.
To overcome this limitation, we introduce the \DPE method. It is inspired from \QRAFT~\cite{qraft} regarding the pseudo oracle and CutQC~\cite{cutqc} to cut the original circuit into smaller subcircuits based on the circuit depth with a specific interval, then calculating the output error with Eq.~\ref{eq:HL_X} for each subcircuit. With the increased circuit depth, the effect of noise also increases~\cite{depthimpact}. Thus, employing multiple depth cuts and computing the output error allows quantifying the noise effect up to a certain depth and provides insights into the temporal evolution of noise's impact on the circuit output.

However, cutting a quantum circuit at various depths has two limitations. First, such cuts may violate the intended logic of a quantum program~\cite{cutqc}, leading to undesirable outcomes. Second, accurate calculation of the circuit error requires knowledge of the ideal output of the circuit at different depths, which is impractical on real quantum computers.
To address these limitations, we leverage the reversible property of a quantum circuit. When a quantum operation transforms a quantum state $|\Psi\rangle$ into a new state $|\Psi^{'}\rangle$ = $U|\Psi\rangle$, the conjugate transpose of $U$, denoted as $U^{\dagger}$, acts as the reverse operation that undoes the effect of $U$~\cite{reversible}. Consequently, when $U^{\dagger}$ is applied to $|\Psi^{'}\rangle$, the original state $|\Psi\rangle$ is restored:
\begin{equation}
U^{\dagger}|\Psi^{'}\rangle = U^{\dagger}(U|\Psi\rangle) = (U^{\dagger}U)|\Psi\rangle = I|\Psi\rangle = |\Psi\rangle 
\end{equation}
Here, $I$ is the identity matrix. With the reversible property, we cut a quantum circuit at a smaller depth and append the inverse of the subcircuit to ensure the quantum logic remains valid. Using the inverse subcircuit, we also get to know the ideal outcome of the quantum circuit, which should always be $|0^{\times{n}}\rangle$ (i.e., all measured qubits should be $|0\rangle$) with a probability of 1 and all other states have a probability of 0.

To calculate \DPE for any quantum circuit, we define depth cut points at the beginning to 1/4\textsuperscript{th} of circuit depth, 1/4\textsuperscript{th} to 1/2\textsuperscript{th} of circuit depth, and 1/2\textsuperscript{th} to 3/4\textsuperscript{th} of circuit depth. We omit the concluding section from 3/4\textsuperscript{th} to the end due to its primary composition of measurement gate operations. Additionally, the inverse of the measurement gate results in no operation on any qubit, which contributes no practical value. We use these ranges to ensure that the total depth of each depth-cut subcircuit, when joined with its inverse, remains less than the total depth of the original full circuit. This condition is essential to prevent correlation weakening due to additional noise from decoherence. Moreover, using these regular interval ranges guarantees that we always calculate a specific number of \DPE features (in this case, three, i.e., \dpeone, \dpetwo, \dpethree for any circuit), making \DPE comparable among different circuits. For example, let's consider a quantum circuit $Q$ with five qubits and a circuit depth of 136. To calculate \DPE for this circuit, we divide $Q$ into three subcircuits: $Q_1$, $Q_2$, and $Q_3$. $Q_1$ represents the subcircuit from depth 0 to 34 (1/4\textsuperscript{th} of 136), $Q_2$ is the subcircuit from depth 35 to 68 (1/2\textsuperscript{th} of 136), and $Q_3$ corresponds to the subcircuit from depth 69 to 102 (3/4\textsuperscript{th} of 136). The \DPE for subcircuits can be calculated using Hellinger distance from Eq.~\ref{eq:HL_X} as $\DPE_n = h(|0^{\times q}\rangle, |Q_nQ_n^{\dagger}\rangle)$, where $q$ stands for the number of measured qubits in the original circuit, $|Q_nQ_n^{\dagger}\rangle$ represents the observed output of the subcircuit appended with its inverse, and $n$ stands for the $n$\textsuperscript{th} subcircuit.

\section{Evaluation and Analysis}
\subsection{Research Question}
We assess the effectiveness of \ourApproach in training ML models to diminish noise errors from quantum circuit outputs by answering the following research questions (RQs):
\begin{itemize}
\item[\textbf{RQ1}] What is the relationship between \ourApproach's feature set and error in circuit output due to noise?
\item[\textbf{RQ2}] How effective is \ourApproach in training ML models for error mitigation, compared with state-of-the-art \QRAFT?
\item[\textbf{RQ3}] Do all features play an important role in mitigating errors from circuit output?
\end{itemize}
\subsection{Experiment Design}

\subsubsection{Benchmarks}
We employ the MQT benchmark~\cite{mqt}, which has a diverse set of quantum circuits tailored for various quantum computers. The circuits in MQT are categorized into two groups: those designed for educational purposes (\textit{learning-level}) and those addressing real-world problems (\textit{application-level}). We selected circuits from both categories that can be executed on all of IBM's quantum computers. In total, we obtained 56 learning-level and six application-level circuits. For application-level circuits, an additional selection criterion was that they must solve real optimization problems. This led to the inclusion of the following circuits in our evaluation.
\begin{inparaenum}[(i)]
\item \textit{Ground State}~(\groundstate): Finds ground state of hydrogen molecules;
\item \textit{Pricing Call}~(\pricingcall): Estimates the fair price of a single European call option using iterative amplitude estimation;
\item \textit{Pricing Put}~(\pricingput): Estimates the fair price of a single European put option using iterative amplitude estimation;
\item \textit{Quantum Approximate Optimization Algorithm}~(\qaoa): Solves a Max-Cut problem instance;
\item \textit{Vehicle Routing}~(\routing): Solves a vehicle routing problem instance; and
\item \textit{Traveling Salesman Problem}~(\tsp): Solves a Traveling Salesman Problem instance.
\end{inparaenum}

MQT benchmark provides the final optimized quantum circuits for all selected problems. For the circuit execution, we used eight industrial IBM quantum computers accessible publicly through the IBM Quantum Cloud platform: \textit{Lagos}, \textit{Nairobi}, \textit{Perth}, \textit{Belem}, \textit{Jakarta}, \textit{Lima}, \textit{Manila}, and \textit{Quito}.

\subsubsection{Machine Learning Models} Error mitigation in a quantum circuit's output is a regression problem. Thus, we selected the most common ML models~\cite{commonml} for regression, i.e., Linear-Regression (LR), Lasso-Regression (Lasso), Ridge-Regression (Ridge), Elastic-Regression (Elastic), Support-Vector Regression (SVR), K-Nearest Neighbour Regression (KNNR), Ensemble-of-Trees-regression (EDT), Light Gradient Boosting Machine (LGBM), Extreme Gradient Boosting (Xgb), and Multilayer-Perceptron (MLP).

\subsubsection{Training and Testing} To generate training data for ML models, we used IBM's quantum simulator in Qiskit Runtime~\cite{qiskit}, with noise models from IBM corresponding to the selected quantum computers. Due to the long waiting queues for getting access to IBM's quantum computers, generating training data on real quantum machines is infeasible. The noise models provided by IBM are constructed using calibration data, approximating the noise experienced in real quantum computers. These models enable classical simulators to produce results with simulated noise. 

For supervised learning, the real probability for each output state is required. Thus, we used an ideal simulator without a noise model to obtain the ground truth for each observed state. However, using an ideal simulator has its limitations, as classical simulators cannot fully simulate complex quantum algorithms. Thus, we selected learning-level circuits that can be simulated on classical computers to generate training data. We hypothesized that the selected features would be generalizable to more complex circuits on real quantum computers. To verify this hypothesis, we used application-level circuits to generate test data by executing them on real quantum computers instead of simulators. 

For the training data, we executed the selected 56 learning-level circuits on noise models of all eight selected quantum computers, resulting in approximately 10k quantum states. We executed the six application-level circuits for the testing data on all eight real quantum computers, yielding 1060 quantum states. For training ML models, we used Optuna~\cite{optuna} for hyperparameter tuning, which uses Bayesian optimization. We opted for the Bayesian optimization to have a fair comparison with \QRAFT since \QRAFT is also trained with Bayesian optimization for hyperparameter tuning. For each trial in Bayesian optimization, the fitness of an ML model was calculated as an average of five-fold cross-validation. For all ML models, we used Mean Square Error as the loss function.

\subsubsection{Metrics}~\label{metrics}
For RQ1, we use Pearson correlation~\cite{pearson} to quantify the relationship between \ourApproach's feature set and the errors in the quantum circuit output caused by noise. For circuit-level features, e.g., depth and width, previous studies (e.g., \cite{depthimpact}) have already demonstrated a positive correlation with quantum noise. Thus, we do not study their correlation. Regarding output-level features, we use the metric \outputstateerror (Eq.~\ref{eq:stateerror}) to assess the impact of noise on a specific output state of a quantum circuit, and the \outputerror (Eq.~\ref{eq:HL_X}) metric to evaluate the overall effect of noise on the circuit's output. We employed Pearson correlation to characterize the relationship between output-level features and the metrics \outputstateerror and \outputerror. Specifically, for \stateweight and \oddsRatio, we computed the correlation with \outputstateerror, as these features are derived for each output state of a circuit. Conversely, for \DPE features, the correlation was determined with \outputerror, given that these features are not calculated for a particular output state but rather for the entire circuit output.

For RQ2, we initially assess the quality of trained ML models using common regression metrics found in the literature~\cite{metric}. These metrics include the Pearson correlation coefficient (\pearsonCoeff), coefficient of determination (\determCoeff), root mean square error (\RMSE), root mean square relative error (\RMSRE), mean absolute percentage error (\MAPE), and mean square error (\TestLoss). For a comparative analysis with \QRAFT, we first identify the best-performing ML model based on the aforementioned regression quality metrics. Subsequently, we compare \ourApproach with \QRAFT using \outputerror~(see Eq.~\ref{eq:HL_X}) as a metric. \outputerror represents the Hellinger distance~(HLD) between two probability distributions. \QRAFT's evaluation mainly used two metrics:
\begin{inparaenum}[(i)]
\item \textit{State Error:} Measured as Mean Square Error (MSE), which in this paper is defined as \TestLoss and used along with other regression metrics.
\item \textit{Program Error:} Initially, in \QRAFT, program error was calculated using Total Variance Distance (TVD). However, we opted for HLD for several reasons.
\end{inparaenum}
HLD is more suitable because it is widely used for comparison with noise~\cite{hellinger3, Hellinger1, Hellinger2}. Also, HLD considers both the difference in probability values and the overall shape of the distribution. This is crucial, as a quantum program may yield low probability outcomes, but as long as the distribution shape aligns with the expected ideal shape, the result is considered correct. In quantum circuits, specific probabilities obtained in an ideal setting may require more shots (i.e., the number of repeated circuit executions to obtain a probability distribution as an output) when subjected to noise. However, if the distribution shape matches the ideal shape under noise, additional shots are unnecessary. TVD does not account for such scenarios, making HLD a more appropriate choice.

For RQ3, we employed the Leave-one-covariate-out (LOCO) method to determine the feature importance for each feature in the proposed feature set. LOCO is a comprehensive method for feature importance assessment~\cite{loco}, involving the exclusion of one feature at a time, retraining the model, and evaluating its performance. In our case, LOCO was utilized to ascertain the impact of a specific feature on \outputerror. To mitigate random bias, we conducted the LOCO process 10 times. For statistical analysis, we employed the Mann-Whitney U~\cite{mann} statistical test and Vargha Delaney \Atwelve~\cite{A12} effect size measure. The statistical tests were conducted on the 10 observations of \outputerror for a given circuit-computer pair without a specific feature, comparing them with 10 observations when utilizing the full feature set. This enables us to assess the importance of each feature for each test circuit-computer pair.

\subsection{Results and Analyses}
\subsubsection{RQ1 -- Relation of \ourApproach's feature set with circuit error}
Effective features must have some relationship with the target value (in our errors in quantum circuit output), allowing ML algorithms to leverage this relationship for predicting the target value. Hence, RQ1 studies such relationship between errors in quantum circuit output and our proposed features. Regarding circuit-level features: circuit depth, width, and counts of single and two-qubit gates, their relationship with noise is well-explored in the literature (e.g., \cite{depthimpact}), which has shown a positive correlation with quantum noise. Therefore, we do not study their correlation in this research question.

For output-level features, instead, we executed the selected ap\-pli\-ca\-tion-level circuits on the selected quantum computers and simulators~\cite{qiskit} and obtained both outputs affected by noise and ideal outputs. To measure the noise effect on a specific output state of a quantum circuit, as well as on overall circuit output, we used the metrics introduced in Section~\ref{metrics}.
Results are shown in Table~\ref{tab:correlation}.
\begin{table}[!tb]
\caption{RQ1 - Results of Pearson correlation analyses for output-level features with \outputstateerror and \outputerror. Columns \stateweight and \oddsRatio denote \textit{State Weight} and \textit{Odds Ratio}; \dpeone, \dpetwo, and \dpethree denote \DPE at various cuts.}
\label{tab:correlation}
\centering
\resizebox{0.45\columnwidth}{!}{%
\begin{tabular}{c|ccccc}
\toprule
\textit{\textbf{Circuits}} & $\stateweight$ & $\oddsRatio$ & $\dpeone$ & $\dpetwo$ & $\dpethree$ \\ 
\midrule
\groundstate & 0.003 & -0.27 & 0.40 & 0.64 & 0.51\\
\pricingcall & -0.44 & -0.16 & 0.49 & 0.94 & 0.95\\
\pricingput & -0.43 & -0.19 & 0.63 & 0.88 & 0.89\\
\qaoa & -0.10 & -0.19 & 0.58 & 0.57 & 0.52\\
\routing & 0.61 & -0.22 & 0.94 & 0.91 & 0.92 \\
\tsp & -0.11 & -0.20 & 0.70 & 0.59 & 0.62\\
\bottomrule
\end{tabular}
}
\end{table}
These correlations explain how the selected features can be used for error mitigation and how they impact the accuracy of quantum circuit outputs.
Column \stateweight in Table~\ref{tab:correlation} shows that for \pricingput and \pricingcall, states with lower weights tend to exhibit higher \outputstateerror due to an overall negative correlation between \stateweight and \outputstateerror.
However, their correlation varies across circuits. For instance, for \routing, a positive correlation was observed. For \groundstate and \qaoa, the correlation is relatively weak. These findings are consistent with \QRAFT~\cite{qraft}, which also employs \stateweight. \QRAFT also highlights that the circuits with a higher number of output states, having larger weights, tend to experience more errors in output states with lower weights. This is because states with higher weights have more qubits in the excited state, making them more prone to relaxation into states with lower weights. This relaxation process accumulates errors in states with lower weights~\cite{qraft}. 

For \oddsRatio and \outputstateerror (see Table~\ref{tab:correlation}), for all circuits, negative correlations were observed, showing that reducing the odds of observing a specific output state leads to an increase in \outputstateerror. This suggests that states with lower odds are more susceptible to noise effect, or they might be noise-induced states (e.g., see Table~\ref{tab:wstate}'s row \textit{Noisy}). Note that the correlation magnitude varies across the circuits, as expected. From columns \dpeone, \dpetwo, and \dpethree, we observe that each \DPE feature exhibits a moderate to strong positive correlation with \outputerror for all circuits implying that \DPE can be used to quantify the noise magnitude affecting a specific circuit output. This makes \DPE a valuable feature for assessing the noise effect.

\begin{tcolorbox}[colback=blue!5!white,colframe=white,breakable]
\textbf{RQ1:} Overall, the Pearson correlation analysis revealed significant correlations for \ourApproach's feature set with errors in quantum circuit output. This suggests that the feature set has the potential to be utilized for mitigating the noise-induced errors in the circuit output. 
\end{tcolorbox}

\subsubsection{RQ2 -- Comparison with \QRAFT}
In this RQ, we train machine learning models using \ourApproach's feature set and compare them with the state-of-the-art \QRAFT. To identify the best-performing regression model to compare with \QRAFT, we used metrics \pearsonCoeff, \determCoeff, \RMSE, \RMSRE, \MAPE, and \TestLoss introduced in Sect.~\ref{metrics}. 

Table~\ref{tab:rq1} presents the results of these metrics for all quantum computers.
\begin{table}[!tb]
\caption{RQ2 -- Performance of ML models on test data for six most commonly used regression performance metrics -- across all eight real quantum computers. Each model with the best performance for a specific metric is in bold}
\label{tab:rq1}
\centering
\resizebox{0.55\columnwidth}{!}{%
\begin{tabular}{l|rrrrrr}
\toprule
\textit{\textbf{Model}} & \pearsonCoeff & \determCoeff & \RMSE & \RMSRE & \MAPE & \TestLoss \\
\midrule
MLP & 0.787 & \textbf{0.605} & \textbf{0.071} & 0.213 & 5.479 & \textbf{0.005} \\
LGBM & 0.792 & 0.567 & 0.074 & 0.227 & 5.655 & \textbf{0.005} \\
XGB & 0.761 & 0.554 & 0.075 & 0.210 & 5.179 & 0.006 \\
EDT & \textbf{0.804} & 0.590 & 0.072 & 0.212 & 5.189 & \textbf{0.005} \\
LR & 0.751 & 0.542 & 0.076 & 0.201 & 5.068 & 0.006 \\
Ridge & 0.751 & 0.542 & 0.076 & 0.201 & 5.068 & 0.006 \\
Lasso & 0.754 & 0.551 & 0.075 & 0.200 & \textbf{5.039} & 0.006 \\
Elastic & 0.754 & 0.551 & 0.075 & 0.200 & \textbf{5.039} & 0.006 \\
SVR & 0.637 & 0.120 & 0.105 & \textbf{0.187} & 5.058 & 0.011 \\
KNNR & 0.710 & 0.485 & 0.080 & 0.227 & 6.272 & 0.006 \\
\bottomrule
\end{tabular}
}
\end{table}
All models exhibit comparable performance across all metrics, indicating that the regression models learned similarly from the selected feature set. 
Higher \pearsonCoeff and \determCoeff values indicate better results, while lower values are preferred for other metrics. Table~\ref{tab:rq1} shows that MLP slightly outperforms the others as it is the best in three metrics; thus, we use it for comparison against \QRAFT.

Table~\ref{tab:rq1_2} presents a comparison between \ourApproach and \QRAFT. The column \textbf{\textit{Obv}} shows the \outputerror without any mitigation, the column \textbf{\textit{M}} represents the \outputerror after applying \ourApproach's mitigation, and the column \textbf{\textit{Q}} shows the \outputerror after \QRAFT's mitigation. The columns \textbf{\textit{\%M}} and \textbf{\textit{\%Q}} demonstrate the percentage improvement from \textbf{\textit{Obv}}. Additionally, the column \textbf{\textit{\%B}} shows the percentage improvement that \ourApproach achieved compared to \QRAFT in column \textbf{\textit{Q}}. In Table~\ref{tab:rq1_2}, positive percentage improvements are highlighted in green, while negative improvements (indicating performance below the compared values) are shown in red.

Table~\ref{tab:rq1_2_circuit} presents the average comparison of \ourApproach's MLP and \QRAFT in terms of \outputerror (Eq.~\ref{eq:HL_X}) for the six application-level quantum circuits and all the selected quantum computers.
\begin{table}[!tb]
\caption{RQ2 -- Comparison of MLP of \ourApproach ($M$) with that of \QRAFT ($Q$), in terms of \outputerror, on IBM's quantum computers and simulators. Column \textit{Obv} shows the averaged values of observed \outputerror without any prediction from \ourApproach's MLP or \QRAFT. Columns \textit{\%M} and \textit{\%Q} show the percentage improvement (given by $\frac{v2-v1}{|v1|}*100$) in error mitigation that \ourApproach's MLP and \QRAFT achieved over \textit{Obv}. \textit{\%B} show the percentage improvement that \ourApproach's MLP achieved over baseline \QRAFT.}.
\label{tab:rq1_2}
\begin{subtable}{\columnwidth}
\caption{Circuit-level (across all selected quantum computers)}
\label{tab:rq1_2_circuit}
\centering
\resizebox{0.7\columnwidth}{!}{%
\begin{tabular}{c|cccccc|cccccc}
\toprule
& \multicolumn{6}{c|}{\textbf{Simulators}} & \multicolumn{6}{c}{\textbf{Real Computers}} \\ \cline{2-13} 
\multirow{-2}{*}{\textbf{Circuit}} & \textit{\textbf{M}} & \textit{\textbf{Q}} & \textit{\textbf{Obv}} & \textit{\textbf{\%M}} & \textit{\textbf{\%Q}} & \textit{\textbf{\%B}} & \textit{\textbf{M}} & \textit{\textbf{Q}} & \textit{\textbf{Obv}} & \textit{\textbf{\%M}} & \textit{\textbf{\%Q}} & \textit{\textbf{\%B}} \\ \midrule
\textit{\groundstate} & 0.20 & 0.36 & 0.35 & \cellcolor[HTML]{ACF495}43.0 & \cellcolor[HTML]{F09C9A}-3.0 & \cellcolor[HTML]{ACF495}44.4 & 0.45 & 0.60 & 0.55 & \cellcolor[HTML]{ACF495}18.0 & \cellcolor[HTML]{F09C9A}-9.0 & \cellcolor[HTML]{ACF495}25.0 \\
\textit{\pricingcall} & 0.32 & 0.34 & 0.54 & \cellcolor[HTML]{ACF495}41.0 & 37.0 & \cellcolor[HTML]{ACF495}6.0 & 0.60 & 0.62 & 0.72 & \cellcolor[HTML]{ACF495}17.0 & 14.0 & \cellcolor[HTML]{ACF495}3.2 \\
\textit{\pricingput} & 0.34 & 0.35 & 0.55 & \cellcolor[HTML]{ACF495}38.0 & 36.0 & \cellcolor[HTML]{ACF495}2.8 & 0.59 & 0.63 & 0.72 & \cellcolor[HTML]{ACF495}18.0 & 12.0 & \cellcolor[HTML]{ACF495}6.3 \\
\textit{\qaoa} & 0.26 & 0.29 & 0.44 & \cellcolor[HTML]{ACF495}41.0 & 34.0 & \cellcolor[HTML]{ACF495}10.3 & 0.57 & 0.48 & 0.64 & 11.0 & \cellcolor[HTML]{ACF495}25.0 & \cellcolor[HTML]{F09C9A}-18.7 \\
\textit{\routing} & 0.05 & 0.12 & 0.23 & \cellcolor[HTML]{ACF495}78.0 & 48.0 & \cellcolor[HTML]{ACF495}58.3 & 0.10 & 0.72 & 0.28 & \cellcolor[HTML]{ACF495}64.0 & \cellcolor[HTML]{F09C9A}-157.0 & \cellcolor[HTML]{ACF495}86.1 \\
\textit{\tsp} & 0.24 & 0.34 & 0.33 & \cellcolor[HTML]{ACF495}27.0 & \cellcolor[HTML]{F09C9A}-3.0 & \cellcolor[HTML]{ACF495}29.4 & 0.36 & 0.66 & 0.43 & \cellcolor[HTML]{ACF495}16.0 & \cellcolor[HTML]{F09C9A}-53.0 & \cellcolor[HTML]{ACF495}45.4 \\ \midrule
\textit{Average} & & & & \textbf{44.6} & 24.8 & \textbf{25.2} & & & & \textbf{24.0} & -28.0 & \textbf{25.0}\\
\bottomrule
\end{tabular}
}
\end{subtable}

\bigskip

\begin{subtable}{\columnwidth}
\caption{Computer-level (across all application-level quantum circuits)}
\label{tab:rq1_2_computer}
\centering
\resizebox{0.7\columnwidth}{!}{%
\begin{tabular}{c|cccccc|cccccc}
\toprule
& \multicolumn{6}{c|}{\textbf{Simulators}} & \multicolumn{6}{c}{\textbf{Real Computers}} \\ \cline{2-13} 
\multirow{-2}{*}{\textbf{Computer}} & \textit{\textbf{M}} & \textit{\textbf{Q}} & \textit{\textbf{Obv}} & \textit{\textbf{\%M}} & \textit{\textbf{\%Q}} & \textit{\textbf{\%B}} & \textit{\textbf{M}} & \textit{\textbf{Q}} & \textit{\textbf{Obv}} & \textit{\textbf{\%M}} & \textit{\textbf{\%Q}} & \textit{\textbf{\%B}} \\ \midrule
\textit{Lagos} & 0.16 & 0.29 & 0.33 & \cellcolor[HTML]{ACF495}52.0 & 12.0 & \cellcolor[HTML]{ACF495}45.0 & 0.44 & 0.60 & 0.54 & \cellcolor[HTML]{ACF495}19.0 & \cellcolor[HTML]{F09C9A}-11.0 & \cellcolor[HTML]{ACF495}27.0 \\
\textit{Nairobi} & 0.24 & 0.34 & 0.41 & \cellcolor[HTML]{ACF495}41.0 & 17.0 & \cellcolor[HTML]{ACF495}29.4 & 0.46 & 0.61 & 0.56 & \cellcolor[HTML]{ACF495}18.0 & \cellcolor[HTML]{F09C9A}-9.0 & \cellcolor[HTML]{ACF495}24.5 \\
\textit{Perth} & 0.21 & 0.33 & 0.40 & \cellcolor[HTML]{ACF495}48.0 & 18.0 & \cellcolor[HTML]{ACF495}36.3 & 0.47 & 0.64 & 0.56 & \cellcolor[HTML]{ACF495}16.0 & \cellcolor[HTML]{F09C9A}-14.0 & \cellcolor[HTML]{ACF495}26.5 \\
\textit{Belem} & 0.43 & 0.29 & 0.56 & 23.0 & \cellcolor[HTML]{ACF495}48.0 & \cellcolor[HTML]{F09C9A}-48.2 & 0.65 & 0.61 & 0.69 & 6.0 & \cellcolor[HTML]{ACF495}12.0 & \cellcolor[HTML]{F09C9A}-6.5 \\
\textit{Jakarta} & 0.16 & 0.26 & 0.34 & \cellcolor[HTML]{ACF495}53.0 & 24.0 & \cellcolor[HTML]{ACF495}38.4 & 0.38 & 0.61 & 0.52 & \cellcolor[HTML]{ACF495}27.0 & \cellcolor[HTML]{F09C9A}-17.0 & \cellcolor[HTML]{ACF495}38.0 \\
\textit{Lima} & 0.23 & 0.34 & 0.41 & \cellcolor[HTML]{ACF495}44.0 & 17.0 & \cellcolor[HTML]{ACF495}32.3 & 0.44 & 0.63 & 0.55 & \cellcolor[HTML]{ACF495}20.0 & \cellcolor[HTML]{F09C9A}-15.0 & \cellcolor[HTML]{ACF495}30.1 \\
\textit{Manila} & 0.25 & 0.26 & 0.41 & \cellcolor[HTML]{ACF495}39.0 & 37.0 & \cellcolor[HTML]{ACF495}3.8 & 0.35 & 0.63 & 0.50 & \cellcolor[HTML]{ACF495}30.0 & \cellcolor[HTML]{F09C9A}-26.0 & \cellcolor[HTML]{ACF495}44.4 \\
\textit{Quito} & 0.20 & 0.28 & 0.39 & \cellcolor[HTML]{ACF495}49.0 & 28.0 & \cellcolor[HTML]{ACF495}28.5 & 0.36 & 0.62 & 0.52 & \cellcolor[HTML]{ACF495}31.0 & \cellcolor[HTML]{F09C9A}-19.0 & \cellcolor[HTML]{ACF495}41.9 \\\midrule
\textit{Average} & & & & \textbf{38.1} & 25.1 & \textbf{21.0} & & & & \textbf{21.0} & -12.3 & \textbf{28.2} \\
\bottomrule
\end{tabular}
}
\end{subtable}
\end{table}
%
The table shows a large difference in noise magnitude between the simulators and real computers. For example, for \groundstate, on average, the observed \outputerror~(\textbf{\textit{Obv}}) on the simulators is 0.35, but on the real computers it's 0.55. This difference is a big error margin.
This difference indicates the need for training data from real computers, which is unfortunately limited due to restricted access. Regarding simulators, the \ourApproach's MLP model demonstrates a substantial improvement in \outputerror (column \textit{\%M}) compared to \QRAFT (column \textit{\%Q}) across all circuits. Though the magnitude of the error mitigation is smaller for real computers, the \ourApproach's MLP model still outperforms \QRAFT, with a total average improvement of 24\%. In contrast, \QRAFT tends to overestimate the noise magnitude, especially in real computer data (highlighted in red). An explanation is that \QRAFT, during feature calculation, doubles the circuit depth to quantify noise, thereby increasing decoherence and cross-talk probabilities, which results in incorrect noise estimation. This is, however, not the case for the \ourApproach's MLP model, as it employs \DPE, which does not need to go beyond the circuit depth. In contrast to the baseline \QRAFT (column \%B), \ourApproach's MLP model demonstrated an average improvement of 25\% for both simulators and real quantum computers. This indicates that the feature set of \ourApproach was effective in enhancing performance, achieving up to a 25\% improvement over the baseline \QRAFT.

Table~\ref{tab:rq1_2_computer} shows the comparison of the \ourApproach's MLP model and \QRAFT regarding \outputerror (Eq.~\ref{eq:HL_X}) for each selected quantum computer across all application-level quantum circuits to study error mitigation capabilities for each specific quantum computer. The table shows that, on average, MLP achieved $38.1\%$ error mitigation compared to \QRAFT's $25.1\%$ on the simulators and $21\%$ as compared to \QRAFT's $-12.3\%$ on real computers. \QRAFT, in general, seems to overestimate noise in real computers with the only exception of \textit{Belem} computer where \QRAFT is better than \ourApproach's MLP. For the rest, \ourApproach's MLP performs better, showing that \ourApproach has the potential to capture noise patterns for most computers. \textit{Belem} being an exception can be caused by two reasons. Either the noise pattern in \textit{Belem} significantly differs from other computers, or the training data is insufficient to generalize across all quantum computers. We will investigate this in the future.

\begin{tcolorbox}[colback=blue!5!white,colframe=white,breakable]
\textbf{RQ2:} Compared to state-of-the-art, \ourApproach is effective in reducing \outputerror on simulators and real quantum computers. Training ML models with \ourApproach on simulator data even outperforms the state-of-the-art method trained on real quantum computers highlighting \ourApproach's potential in capturing noise and motivating its application in training data obtained from real quantum computers. 
\end{tcolorbox}

\subsubsection{RQ3 -- Feature Importance}
We used the Leave-one-covariate-out~(LOCO) feature importance technique introduced in Section~\ref{metrics} to comprehensively assess each feature's impact on the model's performance. To minimize the effect of training variability, we retrained the MLP model with a full feature set ten times. We also executed the LOCO process for each feature ten times. Next, we calculated an average \outputerror for each real computer and application-level circuit pair, resulting in 48 observations for the full feature set and 48 observations for the LOCO process for each feature.
\begin{figure}[!tb]
\centering
\includegraphics[width=0.6\columnwidth]{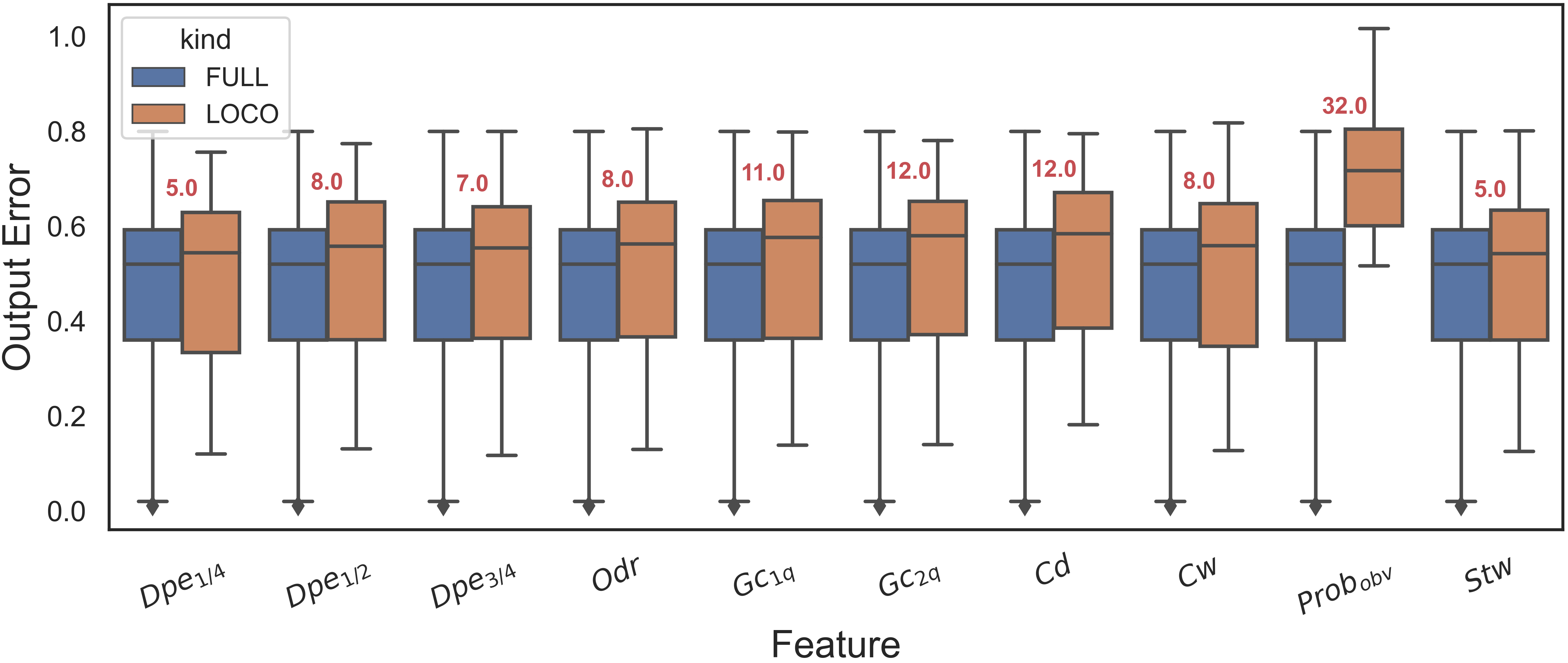}
\centering
\caption{RQ3 -- LOCO feature importance boxplots for the MLP model on test data for real computers and application-level circuits. The text~(red) shows the \%difference between the medians of boxplots with and without a specific feature.}
\label{fig:loco}
\end{figure}
Fig.~\ref{fig:loco} shows that excluding \prob caused a median increase of $32\%$ \outputerror as compared to the full \ourApproach (the blue boxplot). All other features have a median increase of over $5\%$ \outputerror. This shows that all features are important for the MLP model performance. Moreover, we also see slight increases in the variances for the LOCO boxplots as compared to those of the full feature set. This suggests that the full \ourApproach helps to achieve more trustworthy predictions when compared with the models trained after dropping any of the features. 
We also conducted the Mann-Whitney U~\cite{mann} statistical test and Vargha Delaney \Atwelve~\cite{A12} effect size measure. For statistical analysis, we used 10 observations of \outputerror for a circuit-computer pair without a feature and compared them with 10 observations with the full feature set. 
Due to space constraints, we provide a summary of findings; however, full results are provided with the code~\cite{sourcecode}. In summary, out of 48 total circuit-computer pairs, for \dpeone and \dpetwo, there are 71\% pairs, for which removing these features results in a statistically significantly worse \outputerror than keeping them. For \dpethree, the percentage is 62\%, for \oddsRatio is 58\%, for \oneqgate is 75\%, for \twoqgate is 69\%, for \circuitdepth is 60\%, for \circuitwidth is 75\%, for \prob is 88\%, and for \stateweight is 62\%. For all the statistically significant pairs, the \Atwelve effect size was \emph{large} according to the classification of \Atwelve of Vargha et al.~\cite{A12}. 

When looking at all computer-circuit pairs (i.e., a total of 48), we observed that for 48\%, all features were important. For the remaining pairs, we observed that only a subset of features were important. For instance, for \qaoa in the \textit{Lagos} computer, all features' p-values are less than 0.05, with large effect sizes indicating that all features played an important role. On the other hand, for \qaoa on the \textit{Perth} computer, only \dpetwo, \oneqgate, \circuitwidth, and \prob have p-values less than 0.05 with large effect sizes. This insight can be used to recommend specific features for specific computers and circuits, which is an interesting future direction.

\begin{tcolorbox}[colback=blue!5!white,colframe=white,breakable]
\textbf{RQ3:} The \prob feature holds the highest significance and all features contribute significantly to error mitigation.
\end{tcolorbox}

\section{Threats to validity}
\textbf{Construct Validity} pertains to how accurately a measurement assesses the intended theoretical concept. One such threat is associated with the metrics employed to evaluate the effectiveness of \ourApproach. In this work, we used widely accepted regression metrics~\cite{metric}, including \pearsonCoeff, \determCoeff, \RMSE, \RMSRE, \MAPE, and \TestLoss introduced in Section~\ref{metrics} to assess the performance of the machine learning models.

Another concern relates to the choice of metric for comparison with \QRAFT. We opted for the Hellinger Distance (HLD) metric to calculate the output error, in contrast to the Total Variance Distance used by \QRAFT due to: 1) HLD is widely used for evaluating performance in the presence of noise~\cite{hellinger3, Hellinger1, Hellinger2}; 2) Unlike the Total Variance Distance, HLD considers the difference in probability values and the difference in the overall shape of the distribution, which is crucial as a quantum program may produce outcomes with low probabilities, yet the result is deemed correct as long as the distribution shape aligns with the expected ideal shape. In quantum circuits, specific probabilities obtained in an ideal setting may require more shots when subjected to noise. However, if the distribution shape aligns with the ideal shape under noise, additional shots become unnecessary. This makes the HLD more suitable. 

\textbf{Internal Validity} concerns the extent to which experiments can establish a causal relationship between independent and dependent variables. One such threat is about hyperparameters of the ML models. To this end, we used Bayesian optimization for hyperparameter tuning and implemented five-fold cross-validation. This helped mitigate dataset selection bias and ensured a more robust and unbiased evaluation of the models. Another internal validity threat relates to the choice of the depth-cut interval used for calculating the \DPE feature. We opted for 1/4th intervals as they provided a fine segmentation that suited the specific characteristics of the quantum programs under investigation. In many quantum circuits, there are distinctive segments such as state preparation, computational steps, and measurements. The selected intervals closely aligned with these segment divisions in our experiment. For example, in most programs we examined, the last 1/4th of their circuits predominantly consisted of measurement operations. However, we acknowledge that this choice may not be universally optimal. Determining the most suitable intervals requires a separate experiment, and this aspect is part of our future research plans.

\textbf{Conclusion Validity} focuses on the statistical significance of the results derived from an experiment. Using an ML model introduces inherent randomness, meaning that the presented results can exhibit variability. To address this concern, in RQ1, we implemented five-fold cross-validation to minimize the randomness during the ML model training. Regarding RQ2, we assessed ten distinct runs of the complete LOCO experiment to illustrate the median change in performance. The multiple runs of the experiment help dealing with the randomness introduced by ML, providing a more robust and reliable analysis of the outcomes.

\textbf{External Validity} addresses the generalizability of the proposed method to other datasets and domains. One potential challenge in this regard is the number of qubits used in quantum circuits, as the impact of noise significantly increases with a higher number of qubits. Due to limited access to real quantum computers, we used circuits with only 5 to 7 qubits. Note that using simulators with a higher number of qubits does not ensure optimal performance on real quantum computers, as the noise approximation in simulators may differ substantially from that in actual quantum hardware, as evidenced by the \textit{Obv} column in Table~\ref{tab:rq1_2}. The limited number of available circuits for training and testing restricts the generalizability of our findings. To address this limitation, we used the widely adopted benchmark MQT~\cite{mqt} to select circuits that represent commonly used quantum circuits. Our approach involved using simpler circuits for training and more complex, real-world problem-solving circuits for testing, to maintain practical relevance in the noisy QC era.

\section{Discussion}

\subsubsection*{\bf Practical Implications} 
We provide a practical solution to support reliable quantum software development across various stages, including design and implementation, testing, and maintenance. 
During implementation, testing, and maintenance, having a dependable quantum program execution workflow is critical for verifying the developed coding solution and assessing the quantum software's correctness. While quantum software development can initially be done using ideal quantum computer simulators that are noise-free, these simulators become impractical as the complexity of quantum programs increases exponentially with the growing number of qubits~\cite{speed}. For practical quantum programs, developers often have to rely on real quantum computers for program execution. However, this introduces a new level of uncertainty in the output of quantum programs due to the effects of noise.

\begin{figure}[!tb]
\includegraphics[width=0.6\columnwidth]{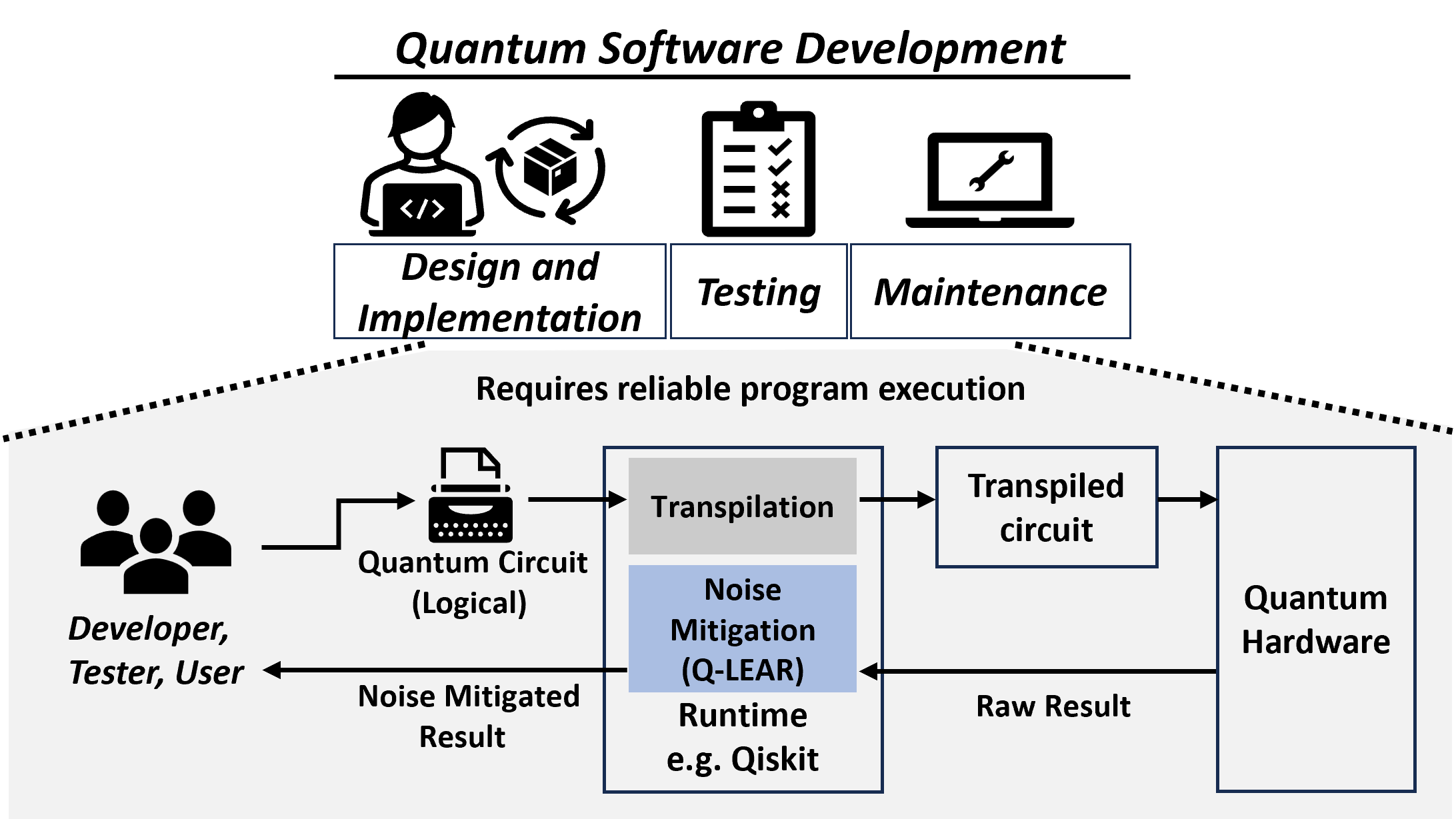}
\centering
\caption{Quantum Program Execution Flow using IBM's Qiskit Framework integrated with \ourApproach}
\label{fig:qse}
\end{figure}

Fig.~\ref{fig:qse} shows the integration of \ourApproach with IBM's Qiskit framework.
By mitigating the noise effect from the noisy output of quantum computer after each program execution, \ourApproach enhances the reliability of outputs obtained from quantum program execution. In our experiments, we demonstrated that \ourApproach can be applied to IBM's quantum computers
as a post-processing approach. In the quantum software development process, when executing quantum software on a quantum computer, \ourApproach empowers quantum software developers to effortlessly mitigate noise from the original outputs produced by quantum software. This capability allows developers to analyze and verify whether the quantum software aligns with their intended behavior, ultimately enhancing the precision and reliability of quantum software development.

\subsubsection*{\bf Integration with Quantum Software Stack} \label{subsec:integrateqiskitstack}
\ourApproach functions as a post-processing module designed to mitigate the effects of noise in the outputs of a quantum program. The process of quantum program execution varies depending on the quantum platform and the framework used for quantum software development. In a typical process, the logical quantum circuit (in our case, written in Qiskit) is first transpiled for the target quantum computer. The execution of the transpiled circuit is then handled by the runtime service provider, such as IBM's Qiskit, and the raw results are obtained from the target quantum hardware. \ourApproach is designed as a separate module that can be integrated with runtime service providers like Qiskit to consume the raw output from quantum hardware and perform noise mitigation. The feature set in \ourApproach comprises generic features that can be calculated from the transpiled version of the quantum circuit and the raw results from the quantum hardware. This design allows \ourApproach to seamlessly integrate with widely used quantum software development frameworks such as Qiskit, as it does not require underlying API calls or logic access from the black-box runtime services and can seamlessly be integrated into the quantum software execution process.

\subsubsection*{\bf Insights on Quantum Noise}
In our experiments, the significant disparity between the classically modeled quantum noise, i.e., noisy simulators, and the noise in real quantum computers is evident, as indicated by Table~\ref{tab:rq1_2} column (\textbf{\textit{Obv}}). Noisy simulators currently serve as a weak approximation of real quantum noise. It is well-established in machine learning that data quality is a crucial component. While training machine learning models for noise mitigation using simulators can yield progress, it has limitations. Our experiments demonstrated that \ourApproach outperforms the state-of-the-art even when trained using data from simulators. However, to achieve substantial improvements in noise mitigation on real quantum computers, there is a pressing need for training data collected from real systems. Presently, wide access to real quantum computers is severely restricted, making it impractical to gather sufficient data for more effective machine learning model training. To support learning-based methods for noise mitigation, there is a need for enhanced infrastructure to support quantum technology. This infrastructure development is a crucial component of the quantum software development roadmap outlined by key companies like IBM~\cite{raodmap}.

\subsubsection*{\bf Generalizability to Other Quantum Computers}
In this work, our experimentation focused on the industrial case study of IBM quantum computers. However, the applicability of \ourApproach is not limited to IBM quantum computers alone. \ourApproach is designed to work with the majority of gate-based quantum computers, including those from IBM, Google, and Rigetti. \ourApproach operates with the transpiled version of a quantum program, and the transpilation process is managed by the vendor's own runtime services such as IBM's Qiskit, Google's Cirq, or Rigetti's Forrest. This design ensures that \ourApproach can work with quantum circuits across multiple quantum computers. Additionally, the feature set employed by \ourApproach is generic and can be computed for all gate-based quantum computers. Currently, the primary differences between quantum computers from different vendors lie in the basis gate set supported and the topology of physical connections. These disparities influence changes in the transpilation process, producing different transpiled quantum circuits for the same logical circuit on different quantum computers. However, \ourApproach's feature set distinguishes between gates based on the number of qubits they act on and calculate features from the transpiled circuits, making it applicable to all current gate-based quantum computers.

Moreover, \ourApproach is a post-processing module independent of runtimes provided by quantum computing vendors. It processes raw outputs from the quantum hardware, which can be served by any gate-based quantum computing runtime such as Qiskit, Cirq, or Forrest. This design choice enhances \ourApproach's generalizability in terms of integration to quantum computers other than IBM. The ML model trained by \ourApproach for IBM computers may not be directly applicable to other quantum computers like Google's Sycamore or Rigetti's Aspen due to variations in quantum noise influenced by the physical and environmental characteristics of each system~\cite{Martina2022}. Retraining the ML model for each quantum computer vendor is necessary, but it's a one-time unavoidable cost due to the inherent nature of quantum noise

\section{Conclusion and Future Work}
To use machine learning (ML) for quantum error mitigation in the current noisy quantum computers, we introduce \ourApproach that utilizes a reliable feature set for training machine learning models. These features are derivable from a quantum circuit and its corresponding output, allowing ML algorithms to mitigate errors in quantum circuit outputs. We evaluated \ourApproach with six application-level quantum circuits on publicly available IBM quantum computers and their corresponding simulators. Our results showed \ourApproach's effectiveness for error mitigation. Our results, in general, show an average improvement of 25\% compared to state-of-the-art on industrial-grade quantum computers and simulators. Our feature importance experiment results show that in general for error mitigation, all features are important. However, for some circuit-computer pairs, the significance of each feature varies. In the future, we will experiment with diverse circuits and investigate the relationship between noise and individual features, particularly how this relationship is influenced by diverse quantum operations across various quantum circuits.

\subsection*{Acknowledgments}
The work is supported by the Qu-Test project (Project \#299827) funded by the Research Council of Norway. S. Ali is also supported by Oslo Metropolitan University's Quantum Hub. Paolo Arcaini is supported by Engineerable AI Techniques for Practical Applications of High-Quality Machine Learning-based Systems Project (Grant Number JPMJMI20B8), JST-Mirai.

\bibliographystyle{IEEEtran}

\begin{thebibliography}{10}
\bibitem{peccost}
D.~Dive, ``With fault tolerance the ultimate goal, error mitigation is the path that gets quantum computing to usefulness,''
\end{thebibliography}

\end{document}